\documentstyle[emulateapj,onecolfloat]{article}

\input epsf
\newcommand{\eal}{\!\!\! &=& \!\!\!}
\newcommand{\al}{\!\!\! && \!\!\!}
\newcommand{\aal}{\!\!\! &\approx& \!\!\!}

\begin{document}

\submitted{Draft \today}

\twocolumn[ 
\title{Reionization Revisited: Secondary CMB Anisotropies and Polarization}
\author{Wayne Hu}
\affil{Institute for Advanced Study, Princeton, NJ 08540}
\begin{abstract}
Secondary CMB anisotropies and polarization provide a laboratory
to study  structure formation in the reionized epoch.
We consider the kinetic Sunyaev-Zel'dovich effect
from mildly nonlinear large-scale structure 
and show that it is a natural extension of the perturbative
Vishniac effect.  If the gas traces the dark matter to overdensities of
order 10, as expected from simulations, this effect is at least comparable to the Vishniac
effect at arcminute scales.  On smaller scales, it may be
used to study the thermal history-dependent clustering of the gas.
Polarization is generated through Thomson scattering of primordial quadrupole anisotropies,
kinetic (second order
Doppler) quadrupole anisotropies and intrinsic scattering quadrupole
anisotropies.  Small scale polarization results from the density and ionization modulation
of these sources.  These effects generically produce
comparable $E$ and $B$-parity polarization, but of negligible amplitude
($10^{-3}-10^{-2}$ $\mu K$) in adiabatic CDM models.   
However, the primordial and kinetic quadrupoles are observationally comparable
today so that a null detection of $B$-polarization would set constraints
on the evolution and coherence of the velocity field.  
Conversely, a detection of a cosmological $B$-polarization 
even at large angles does not necessarily imply the presence of
gravity waves or vorticity. 
For these calculations, we develop an all-sky generalization of 
the Limber equation that allows for an arbitrary local angular dependence 
of the source for both scalar and symmetric trace-free tensor 
fields on the sky.
\end{abstract}
]

\section{Introduction}

With the rapid improvement in the sensitivity of cosmic microwave background (CMB) experiments 
in recent years, it becomes increasingly important to address small contributions to the anisotropy
and polarization at the $\mu $K level and below.  While the primary anisotropies and polarization
from the epoch of recombination are thought to be understood at this level theoretically, secondary anisotropies
from reionization are not.  In this work, we revisit the generation of these secondary anisotropies,
uncovering a potentially important contribution from the nonlinear regime, and explicitly calculate the 
small secondary polarization signal.   We consider only those contributions from
reionization that are true, frequency-independent, temperature distortions.   Isolating these
contributions observationally from potentially larger but frequency-depen\-dent foregrounds will
be a great challenge but one that lies beyond the scope of this paper (see e.g. \cite{BouGis99} 1999;
\cite{Tegetal99} 1999).

Even the epoch of reionization itself cannot be accurately predicted in the context of an otherwise
precisely defined theory of structure formation.  The efficiency with which ionizing photons are
produced and escape from the first baryonic objects in the universe will likely remain uncertain
even with substantial advances in numerical simulations (see e.g. \cite{AbeNorMad98} 1998).
In the adiabatic cold dark matter (CDM) class of
models, reionization is expected to occur in the range $8 \la z_i \la 20$ but with large uncertainties
(see e.g. \cite{HaiKno99} 1999 for a recent review).
Observationally, reionization must be essentially complete by 
$z \sim 5$ due to the absence of the
Gunn-Peterson effect in quasar absorption spectra (\cite{GunPet65} 1965).  For models with a roughly
scale-invariant spectrum of initial fluctuations, detections of degree-scale anisotropies imply optically
thin conditions or $z_i \la 50$ (\cite{GriBarLid99} 1999).  The large angle polarization of the 
CMB can in principle yield precise constraints on the reionization epoch but has not been detected to date.

Even if reionization took place around $z=5$, secondary anisotropies from the \cite{Vis87} (1987)
effect should produce $\mu $K anisotropies in the arcminute regime
(\cite{HuWhi96} 1996).  The reason
is that this is a second order effect whose amplitude per logarithmic interval in $(1+z)$ is
constant, i.e. the contributions from below $z=5$ are similar in magnitude to those in the
range $5 < z \la 30$.   In the optically thin adiabatic CDM models, at least 
$\sim 1/2$ of the 
Vishniac effect comes from redshifts 
$z \la 5$. Given these rather low redshifts of formation, it is interesting to consider whether nonlinear
effects can further enhance the anisotropy.  
After developing general techniques for calculating secondary anisotropies 
and polarization in \S \ref{sec:general}, 
we consider contributions from the kinetic Sunyaev-Zel'dovich
effect from large scale structure in the mildly nonlinear regime in 
\S \ref{sec:vishniactemp},  
and show that it is the natural 
nonlinear extension of the Vishniac effect. 
On small scales, both arise from the density modulation of
the Doppler effect from large-scale bulk flows.  

We then turn to secondary polarization.  
Back-of-the-envelope estimates in these optically thin conditions immediately place the secondary 
polarization signal orders of magnitude below
the secondary anisotropies, except for the well-studied large-angle polarization from reionization.
Nevertheless given the great potential of precision polarization measurements
for studying gravity waves (\cite{KamKosSte97} 1997; \cite{ZalSel97} 1997),
an explicit calculation is of interest. 
In \S \ref{sec:homopol}, we treat the polarization that results from second-order Doppler shifts 
due to bulk flows and separate the contributions in the two parity modes.
This calculation may also be of interest to studies of CMB polarization in galaxy clusters.
In \S \ref{sec:vishniacpol}, we consider the analogue of the Vishniac effect for polarization:
the density modulation of the kinetic and primordial polarization sources as 
well as the generation of polarization through rescattering of Vishniac temperature anisotropies.   
Finally in \S \ref{sec:patchy}
we show that the same relationship between anisotropies and polarization for density-modulated signals
holds for ionization-fraction modulated signals, as relevant for the brief epoch of inhomogeneous 
ionization expected at the onset of reionization.  

The basis for all of these calculations is presented in the Appendix.  There we generalize 
the \cite{Lim54} (1954) approximation for sources with arbitrary local angular dependence
and fields on the sky that are either scalar 
or tensor in nature.  These formulae also 
extend the flat sky Limber approach to the full sky and may be useful in other cosmological studies.

\section{General Considerations}
\label{sec:general}

We review the relevant 
properties and parameters of the adiabatic CDM 
scenario for structure formation in \S\ref{sec:adiabatic}. 
In \S \ref{sec:secondaryT} and \S \ref{sec:secondaryP}, we discuss the general principles involved in calculating
secondary CMB anisotropies and polarization 
based on formalism developed in the Appendix.  

\subsection{Adiabatic CDM Model}
\label{sec:adiabatic}

We work in the context of the adiabatic cold dark matter
(CDM) family of models where structure forms through the gravitational
instability of the CDM in a background
Friedmann-Robertson-Walker metric.  In units of the critical density 
$3H_0^2/8\pi G$, where $H_0=100h$ km s$^{-1}$ Mpc$^{-1}$ is the
Hubble parameter today, 
the contribution of each component is denoted $\Omega_i$,
$i=c$ for the CDM, $b$ for the baryons, $\Lambda$ for the cosmological
constant.
It is convenient to define
the auxiliary quantities $\Omega_m=\Omega_c+\Omega_b$ and  $\Omega_K=1-\sum_i \Omega_i$, which represent
the matter density and 
the contribution of spatial curvature to the expansion rate respectively. The expansion 
rate then becomes
\begin{equation}
H^2 = H_0^2 \left[ \Omega_m(1+z)^3 + \Omega_K (1+z)^2 
		+\Omega_\Lambda \right]\,.
\end{equation} 

Convenient measures of distance and time include the conformal distance (or lookback
time) from the observer at redshift $z=0$ in units of the Hubble distance today $H_0^{-1}
=2997.9h^{-1} $Mpc,
\begin{equation}
D(z) = \int_0^z {H_0 \over H(z')} dz' \,,
\end{equation}
and the analogous angular diameter distance
\begin{equation}
D_A = \Omega_K^{-1/2} \sinh (\Omega_K^{1/2} D)\,.
\end{equation}
Note that as $\Omega_K \rightarrow 0$, $D_A \rightarrow D$.

The adiabatic CDM model possesses a 
power spectrum
of fluctuations in the gravitational potential $\Phi$
\begin{equation}
\Delta_{\Phi}^{2} = {k^3 \over 2\pi^2} P_{\Phi} = A\left({k \over H_0}\right)^{n-1}T^2(k) \,,
\end{equation}
where the transfer function $T(k)=1$ for scales much larger than the horizon at matter-radiation
equality.  We employ the CMBFast code 
(\cite{SelZal96} 1996)
to determine $T(k)$ at intermediate scales and extend
it to small scales using the fitting formulae of \cite{EisHu99} (1999).
Throughout this paper, the notation $\Delta^2_S$ will always represent the logarithmic
power spectrum of the field ``$S$'' and should be assumed to be time-dependent
even where that argument is suppressed.  The rms of the field is defined as
\begin{equation}
S_{\rm rms}^2 = \int {dk \over k}\Delta_S^2 \,.
\end{equation}  

The cosmological Poisson equation
relates the power spectra of the potential and density perturbations $\delta$
\begin{equation}
\Delta_\Phi^2 = {9 \over 4} \left( {H_0 \over k} \right)^4
\left(1 + 3{H_0^2 \over k^2}\Omega_K \right)^{-2} \Omega_m^2 (1+z)^{2}\Delta_\delta^2\,,
\end{equation}
and gives the relationship between their relative normalization
\begin{equation}
A = {9 \over 4} 
\left(1+ 3 {H_0^2 \over k^2} \Omega_K \right)^{-2} 
\Omega_m^2 (1+z)^{2} G^2 \delta_H^2  \,.
\label{eqn:normalization}
\end{equation}
Here $\delta_H$ is the amplitude of present-day density fluctuations at the Hubble scale;
we adopt the COBE normalization for $\delta_H$ (\cite{BunWhi97} 1997).
$G(z)$ is the growth rate of linear density perturbations
$\delta(z)=G(z)\delta(0)$.  Expressions and approximations for this growth function can be found in 
the literature (e.g. \cite{Pee80} 1980; \cite{CarPreTur92} 1992) 
and when the energy density in radiation can be neglected
\begin{equation}
G(z) \propto {H(z) \over H_0} \int_z^\infty dz' (1+z') \left( {H_0 \over H(z')} \right)^3\,.
\end{equation}
For the matter dominated regime where $H \propto (1+z)^{3/2}$, 
$G \propto (1+z)^{-1}$ and $A=$ const.
Likewise, the continuity equation relates the density and velocity 
power spectra
\begin{eqnarray}
\Delta_{v}^2 = \left( {\dot G \over G}{H_0 \over k} \right)^2 
{\Delta_{\delta}^2 } \,.
\label{eqn:velocitypower}
\end{eqnarray}

For this type of fluctuation spectra and growth rate, 
reionization of the universe
is expected to occur rather late $z_i \la 50$ such that the reionized
media is optically thin to Thomson scattering of CMB photons 
$\tau \la 1$.
The probability of last
scattering within $d D$ of $D$ (the visibility function) is 
\begin{equation}
g =  \dot \tau e^{-\tau} = X \tau_H (1+z)^2 e^{-\tau}\,.
\end{equation}
Here
dots represent derivatives with respect to $D$, 
$X$ is the ionization fraction, and
\begin{equation}
\tau_H = 0.0691 (1-Y_p)\Omega_b h\,,
\end{equation}
is the optical depth to Thomson
scattering to the Hubble distance today, assuming full
hydrogen ionization,
where $Y_p$ is the primordial helium fraction.
Note that the ionization 
fraction can exceed unity:
$X=(1-3Y_p/4)/(1-Y_p)$  for singly ionized helium,
$X=(1-Y_p/2)/(1-Y_p)$ for fully ionized helium.  We 
assume that $X=1$ in the reionized epoch such that 
\begin{equation}
\tau = {2 \over 3} {\tau_H \over \Omega_m^2} [2-3\Omega_m+\sqrt{1+\Omega_m z}(\Omega_m z + 3\Omega_m -2)]\,,
\end{equation}
for $\Omega_\Lambda=0$ and
\begin{equation}
\tau = {2 \over 3} {\tau_H \over \Omega_m} [\sqrt{1-\Omega_m+\Omega_m(1+z)^3}-1]
\end{equation}
for $\Omega_K=0$.

\begin{figure}[t]
\centerline{\epsfxsize=3.5truein\epsffile{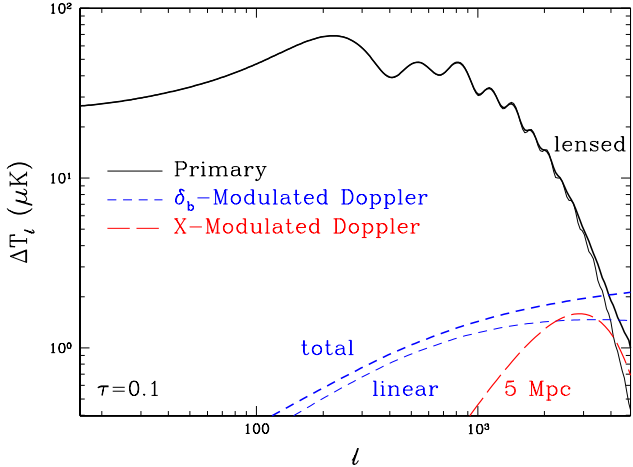}}
\caption{Temperature anisotropies for the fiducial $\Lambda$CDM model with $\tau=0.1$ ($z_i=13$). 
The secondary density ($\delta_b$) modulated signal has been calculated under the assumption that
the gas traces the dark matter.  The ionization-modulated signal assumes patches of $5$ Mpc 
comoving size and duration of patchiness $\delta z_i/(1+z_i)=0.25$.
}
\label{fig:temp}
\end{figure}

\begin{figure}[t]
\centerline{\epsfxsize=3.5truein\epsffile{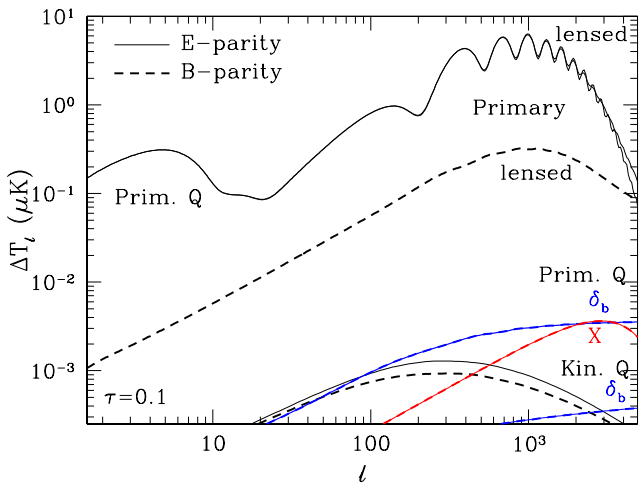}}
\caption{Polarization for the fiducial $\Lambda$CDM model with $\tau=0.1$ 
separated
into $E$ (solid lines) and $B$ (dashed lines) contributions.
Secondary anisotropies from the primordial quadrupole are labeled (Prim. $Q$):
(upper) homogeneous scattering; (lower) density ($\delta_b$) and ionization ($X$) modulated scattering
following Fig.~\protect\ref{fig:temp}.  For the kinematic quadrupole, the homogeneous and density
modulated signals are shown; the ionization modulated and intrinsic quadrupole signals falls below this range.   
Note that the $B$-parity polarization induced by gravitational lensing
is much larger than any of these secondary $B$ signals.
}
\label{fig:pol}
\end{figure}

Although we maintain generality in all derivations, we 
illustrate our results with a $\Lambda$CDM model.  
The parameters for this model
are $\Omega_c=0.30$, $\Omega_b=0.05$, $\Omega_\Lambda=0.65$, $h=0.65$, 
$Y_p = 0.24$, $n=1$, and $\delta_H=4.2 \times 10^{-5}$.
This model has mass fluctuations on the $8 h$ Mpc$^{-1}$
scale in accord with the abundance of galaxy clusters
$\sigma_8=0.86$.  A reasonable value here
is important since many of the effects discussed below are nonlinearly
dependent on the amplitude of mass fluctuations at or below this scale.
We consider reionization redshifts in the range $5 \la z_i \la 30$ or $0.025 \la \tau \la 0.3$.  
For reference, the primary anisotropies and polarization for a model with $\tau=0.1$ ($z_i=13$) is
shown in Figs.~\ref{fig:temp} and \ref{fig:pol} respectively

%for the CDM, $\Omega_b (=0.05)$ for the baryons, $\Omega_\Lambda (=0.65)$
%for the cosmological constant.  Values in parentheses represent the model
%which we will use to illustrate the results. 

%\begin{eqnarray}
%\delta T_\ell^2 \eal {\ell (\ell+1) \over 2\pi} C_\ell^{\Theta\Theta}  T_{\rm cmb}^2 
%\nonumber\\
%\delta P_\ell^2 \eal {\ell (\ell+1) \over 2\pi} [ C_\ell^{EE} + C_\ell^{BB} ] T_{\rm cmb}^2 
%\end{eqnarray}

%\section{Modulated Thomson Scattering}
%\label{sec:modulated}

\subsection{Secondary Anisotropies}
\label{sec:secondaryT}

The motion of the scatterers imprints a temperature fluctuation
on the CMB through the Doppler effect
\begin{equation}
\Theta(\hat{\bf n}) = \int dD\, g(D) {\hat {\bf n}} \cdot {\bf v}_b({\bf x}) \,,
\end{equation}
where 
$\hat {\bf n}$ is 
the direction on the sky, and ${\bf v}_b$ is the velocity field
of the baryons and is evaluated along the line of sight ${\bf x}=
D \hat{\bf n}$.  

Let us first consider the case where the visibility $g$ is dependent on
time only.  The source
${\hat {\bf n}} \cdot {\bf v}_b$ is 
a field that depends both on direction and on spatial position.
It is useful to decompose the Fourier transform of the velocity
field into a scalar and two vector (vorticity) modes
\begin{equation}
{\bf v}_b({\bf k}) = -i v_b^{(0)} {\bf e}_3 
		   + \sum_{m=\pm 1} v_b^{(m)} {{\bf e}_2 \mp i{\bf e}_1 
			\over \sqrt{2}} \,,
\label{eqn:velocitydecomp}
\end{equation}	
such that
\begin{equation}
\hat{\bf n} \cdot {\bf v}_b({\bf k}) = - i \sqrt{4 \pi \over 3}
	\sum_{m=-1}^1 v_b^{(m)} Y_1^{(m)}(\hat{\bf n})\,,
\end{equation}
where ${\bf e}_3 \parallel \hat{\bf k}$.  As shown in the Appendix,
the angular power spectrum resulting from the weighted projection of 
a source with a local dipolar angular dependence is
(see equation~[\ref{eqn:limberform}]),
\begin{eqnarray}
C_\ell^{\Theta\Theta} \eal {\pi^2 \over \ell^3} 
	\int dD\, g^2 D_A 
	\sum_{m=\pm 1}  \Delta_{v_b}^{2\,(m)}  \,,
\label{eqn:dopplervorticity}
\end{eqnarray}
for $\ell \gg 1$.
The power per logarithmic interval in the velocity field
$\Delta_{v_b}^{2\,(m)}$ is 
evaluated at the wavenumber that projects onto the angular
scale $\ell$ at $D$, 
\begin{equation} 
k=H_0 {\ell \over D_A}\,.
\label{eqn:kproj}
\end{equation}  
The implicit assumption here is that the visibility-weighted source
$g(D)v_b(k,D)$ is slowly-varying
across a wavelength of the perturbation.

We will often use 
the power per logarithmic interval in $\mu$K$^2$ or $n$K$^2=(10^{-3} \mu$K)$^2$ 
\begin{equation}
\Delta T_\ell^2 = {\ell(\ell+1) \over 2\pi} C_\ell^{\Theta\Theta} 
	T_{\rm CMB}^2 \,,
\label{eqn:logpower}
\end{equation}
where $T_{\rm CMB} = 2.728 \times 10^6$ $\mu$K.
The rule of thumb for $\Delta T_\ell^2$ is that it is of order the
logarithmic power spectrum of the source (at cosmological distances 
$D_A \sim 1$) divided by the multipole $\ell$.  This factor of
$\ell$ comes from the loss of modes parallel to the line of sight due to 
crest-trough cancellation of the contributions.

Notice however that the $m=0$ potential flow component drops out of 
the final expression and violates this rule. 
In a potential flow 
waves
perpendicular to the line of sight lack a velocity component 
parallel to the line of sight; consequently
there is no Doppler effect to leading order (\cite{Kai84} 1984).
 \cite{OstVis86} (1986)
pointed out that the same is
not true for vortical flows since waves that run perpendicular to the line of sight have
velocities parallel to the line of sight.  Since flows
in the linear regime are potential, the leading order Doppler effect
is nonlinear in the perturbations at small scale. 

It is not necessary for the flows themselves to possess
vorticity.
Since it is the visibility-weighted
velocity field $g {\bf v}_b$ that is the real source,
spatial modulations in the visibility 
create an effective velocity field
\begin{equation}
{\bf v}_b (1 + \delta g/g) \equiv 
{\bf v}_b + {\bf v}_g \,,
\label{eqn:vmodulated}
\end{equation}
which can be used in place of ${\bf v}_b$ in
equation~(\ref{eqn:dopplervorticity}). 
Spatial variations in the
free electron density modulate the visibility and themselves 
may be caused by fluctuations in the
net baryon density (see \S \ref{sec:vishniactemp}) or the ionization fraction
(see \S \ref{sec:patchy}).

\subsection{Secondary Polarization}
\label{sec:secondaryP}

Thomson scattering of radiation with 
a quadrupole anisotropy 
\begin{equation}
Q^{(m)}({\bf x}) = -\int d\Omega \, { Y_2^{m*} ( \hat{\bf n}) \over \sqrt{4\pi}}
	\Theta({\bf x},\hat{\bf n})
\end{equation}
generates
linear polarization in the CMB.
In terms of the Stokes $q$ and $u$
parameters
\begin{eqnarray}
[q\pm i u](\hat{\bf n}) \eal{ \sqrt{ 24\pi} \over 10 }\int dD\, g
	\sum_{m=-2}^2 
	Q^{(m)}
%\nonumber\\
%	\al \quad \times 
{}_{\pm 2} Y_2^{m}(\hat{\bf n}) \,,
\end{eqnarray}
where $Q^{(m)}$ are the 5 quadrupole moments of the temperature 
fluctuation\footnote{$Q^{(m)}= \Theta_2^{(m)}/\sqrt{5}$ in the notation of
\cite{HuWhi97} (1997), see their eqn. [77].} and
${}_{\pm 2} Y_\ell^m$ are the spin-2 spherical harmonics (\cite{Goletal67} 
1967).  

Some subtleties arises when defining the Fourier transform of the quadrupole
source since the orientation of the coordinate system
enters into its definition.
It is convenient to chose the reference frame so that $\hat{\bf e}_3 \parallel {\bf k}$,
and this choice will be understood unless otherwise specified.  The 
consequence is an ambiguity in combining 
contributions from different ${k}$-modes but these can be removed by 
defining the coordinate independent 
components of the polarization (\cite{ZalSel97} 1997;
\cite{KamKosSte97} 1997),
\begin{eqnarray}
[q\pm i u](\hat{\bf n}) \eal \sum_\ell \sum_{m=-2}^2 [E_\ell^{(m)} \pm i B_\ell^{(m)}]
{}_{\pm 2} Y_2^{m}(\hat{\bf n}) \,.
%\nonumber\\
%	\al \quad \times 
\end{eqnarray}

Since polarization is a spin-2 field and its source has a local
quadrupole angular dependence, its power spectrum is given by
equation~(\ref{eqn:limberform}) with $s=\pm 2$ and $j=2$.
The power spectra of the $E$ and $B$ parity states
are then
\begin{eqnarray}
C_\ell^{EE} \eal {3 \pi^2 \over 10\ell^3} \int dD\, g^2 D_A 
	\left( {3 \over 4} \Delta_{Q}^{2\,(0)} +
	       {1 \over 8}\sum_{m=\pm 2} \Delta_{Q}^{2\,(m)}  \right) \,,
\nonumber\\
C_\ell^{BB} \eal {3 \pi^2 \over 10\ell^3} \int dD\, g^2 D_A 
	{1 \over 2} \sum_{m=\pm 1} \Delta_{Q}^{2\,(m)} \,,
\label{eqn:eebb}
\end{eqnarray} 
again evaluated with the projection equation (\ref{eqn:kproj}) and
a slowly-varying source.
We define the logarithmic power spectrum of the polarization 
in the same way as for the 
anisotropies (see equation~[\ref{eqn:logpower}]).  
It is likewise down by a factor of $\ell$ from
the power spectrum of the source except when the angular dependence of
the source reduces it further as in the case of $B$-parity polarization 
from $m=0,\pm 2$ and $E$-parity polarization from $m=\pm 1$.  

As pointed out by \cite{Kai84} (1984) and examined quantitatively
by \cite{Efs88} (1988), the polarization, even more than the temperature
anisotropy, 
is suppressed by cancellation at small scales. 
The reason is simply that the source of the polarization {\it is}
the temperature anisotropy (quadrupole moment) and thus both the source
and its 
contributions along the line of sight are suppressed. 

The dominant source at small angles is again the visibility modulation
of the large scale quadrupole anisotropy
\begin{equation}
Q^{(m)} \rightarrow  Q^{(m)} (1 + \delta g/g) 
\equiv Q^{(m)} + Q_g^{(m)}  \,.
\label{eqn:qmodulated}
\end{equation}
We will consider quadrupole sources $Q^{(m)}$ in \S \ref{sec:homopol},
the effect of density modulation in 
\S \ref{sec:vishniacpol}, and that of ionization modulation in \S \ref{sec:patchy}.

\section{Density-Modulated Anisotropies}

In this section we consider the temperature anisotropies that arise from the modulation of
the Doppler effect due to density inhomogeneities.  We begin with a review of the
second-order effect from linear density fluctuations (\S \ref{sec:linearV}) 
and then generalize the calculation
to include small-scale nonlinearities in the density field (\S \ref{sec:nonlinearV}). 

\label{sec:vishniactemp}

\subsection{Linear Fluctuations: Vishniac Effect}
\label{sec:linearV}

Spatial variations in the opacity due to density perturbations
in the baryons modulates the visibility: $\delta g/g = \delta_b$.  When $\delta_b$ is 
in the linear regime, the result is called the \cite{Vis87} (1987)
effect. 
In this section, we rederive the Vishniac effect in a manner that
will make its nonlinear generalization obvious.

The Fourier transform of ${\bf v}_g$ is a convolution
of the linear velocity and density fields,
\begin{equation}
{\bf v}_g({\bf k})=
\int {d^3 k_1 \over (2\pi)^3 }
{\bf v}_b({\bf k}_1) \delta_b (k_2)  \,,
\end{equation}
where here and througout
\begin{equation}
{\bf k}_2 = {\bf k} - {\bf k}_1 \,.
\end{equation}
The two vortical components
to the velocity field are given by the projections  
(see equation~[\ref{eqn:velocitydecomp}])
\begin{eqnarray}
v_g^{(\pm 1)}({\bf k}) \eal
{\hat{\bf e}_2 \pm i \hat{\bf e}_1 \over \sqrt{2}} \cdot 
{\bf v}_g \,, 
\label{eqn:vishsource}
\\
	\eal
\sqrt{4\pi \over 3} \int {d^3 k_1 \over (2\pi)^3 }
v_b(k_1) \delta_b(k_2)Y_1^{\pm 1*}(\hat{\bf k}_1)  \,,
\nonumber
\end{eqnarray}
in the basis where ${\bf e}_3 \parallel {\bf k}$.  We have
assumed that the underlying velocity field is a potential
flow ${\bf v}_b({\bf k}) = -i v_b(k)\hat{\bf k}$.
 
The power spectrum then becomes
\begin{eqnarray}
\Delta_{v_g}^{2\, (1)} 
={1 \over 3}
\int {dk_1 \over k_1} d \Omega \,{k^3 \over k_2^3} 
 |Y_1^{1}({\bf k}_1)|^2 [A + B]\,,
\label{eqn:velocitycorr}
\end{eqnarray}
and likewise for the $m=-1$ component.
The two contributions are from the velocity-velocity, density-density
power spectra
\begin{equation}
A = \Delta^2_{v_b}(k_1) \Delta^2_{\delta_b}(k_2) \,,
\label{eqn:vvdd}
\end{equation}
and the velocity-density cross correlation power spectra
\begin{equation}
B =  - {k_1 \over k_2}
	\Delta^2_{v_b \delta_b}(k_1)\Delta^2_{v_b \delta_b}(k_2)
\,,
\label{eqn:vdvd}
\end{equation}
where we have used the relation
\begin{equation}
Y_1^1({\bf k}_2) = -{k_1  \over k_2} Y_1^1({\bf k}_1)\,.
\end{equation}
Substituting equation~(\ref{eqn:velocitycorr}) and 
(\ref{eqn:dopplervorticity}) after reexpressing the velocity power
spectrum with the density power spectrum using equation~(\ref{eqn:velocitypower}),
we obtain
\begin{eqnarray}
C_\ell^{\Theta\Theta} \eal {\pi^2 \over 2\ell^5 } \int
{dD}\, D_A^3 
\left( g {\dot G \over G} \right)^2 
%\nonumber\\ 
%\al\quad \times
\Delta_{\delta_b}^4 I_V \,,
\label{eqn:vishmode}
\end{eqnarray}
where we have used the fact that $m = \pm 1$ contributes equally.
The mode coupling integral is
\begin{eqnarray}
I_V \eal \int_0^\infty dy_1 \int_{-1}^1 d\mu {(1-\mu^2)(1-2\mu y_1) 
	\over y_1^3  y_2^5 } 
%\al \quad \times
{   \Delta^2_{\delta_b}(k y_1) \over
    \Delta^2_{\delta_b}(k) }
{   \Delta^2_{\delta_b}(k y_2) \over
    \Delta^2_{\delta_b}(k) }
	\,.
\nonumber\\
\label{eqn:modecoupling}
\end{eqnarray}
Here and throughout
\begin{eqnarray}
\mu \eal \hat{\bf k} \cdot \hat{\bf k}_1  \,,
\nonumber\\
y_1 \eal k_1/k \,,
\nonumber\\
y_2 \eal k_2/k = \sqrt{1 - 2\mu y_1 + y_1^2} \,.
\end{eqnarray}

The form of the
mode coupling integral in 
equation (\ref{eqn:modecoupling}) is 1/2 that of \cite{Vis87} (1987) equation (2.13),
which resolves the apparent discrepancy with
\cite{DodJub95} (1995) raised
by \cite{JafKam98} (1998).
As pointed out by \cite{DodJub95} (1995), this form is easier 
to evaluate numerically.  Additionally, for our purposes, it
better brings out the small-scale limit.

\begin{figure}[t]
\centerline{\epsfxsize=3.5truein\epsffile{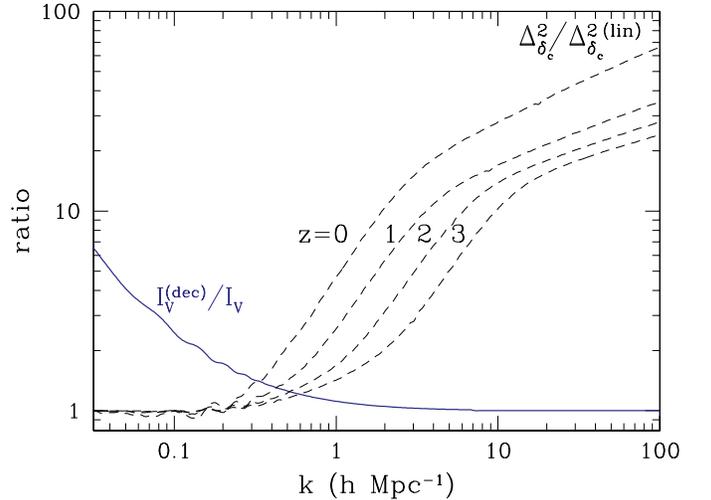}}
\caption{Velocity-density decoupling and density nonlinearities.  For the fiducial $\Lambda$CDM
model the velocity and density fields decorrelate in the Vishniac calculation ($I_V^{\rm dec}/I_V \approx 1$)
before the onset of nonlinearity in the density field, especially at high redshift where most of the 
scattering occurs. 
}
\label{fig:I}
\end{figure}

For wavelengths that are much smaller than the coherence scale of 
the velocity field, these expressions simplify considerably.  
The density-velocity cross correlation (\ref{eqn:vdvd}) vanishes
and the remaining term (\ref{eqn:vvdd}) can be evaluated under
the approximation $k_2 = |{\bf k}-{\bf k}_1| \approx k$,
\begin{equation}
\Delta_{v_g}^{2\, (\pm 1)} =
	{1 \over 3} \Delta^2_{\delta_b} v_{\rm rms}^2 \,,
\label{eqn:vdecoherent}
\end{equation}
where we have used the orthogonality of spherical harmonics
$\int d\Omega\, Y_{\ell}^{m*} Y_{\ell'}^{m'} = \delta_{\ell,\ell'}
\delta_{m,m'}$ and 
\begin{equation}
v_{\rm rms}^2  = \int {dk \over k} \Delta_{v_b}^2 \,.
\end{equation}  
Note that $v_{\rm rms}^2$ is still a function of $D$.

Equation~(\ref{eqn:vishincoherent}) has a simple interpretation. 
At small scales, the effect arises from a density modulation
of a uniform bulk velocity from larger scales whose
rms in each component is 1/3 of the total.

Substituting the velocity power spectrum relation 
(\ref{eqn:vdecoherent}) 
into equation~(\ref{eqn:velocitycorr}) yields
\begin{eqnarray}
C_\ell^{\Theta\Theta} \eal {2\pi^2 \over 3\ell^3} \int
{dD}\, D_A
g^2 
\Delta_{\delta_b}^2 v_{\rm rms}^2 \,.
\label{eqn:vishincoherent}
\end{eqnarray}
This equation may alternately be derived from
equation (\ref{eqn:vishmode}) in the limit that the integral
gets its contribution from $y_1 \ll 1$,
\begin{eqnarray}
I_V^{\rm (dec)} \eal {4 \over 3} \int_0^\infty {dy_1 \over y_1^3}
{   \Delta^2_{\delta_b}(k y_1,\eta) \over
    \Delta^2_{\delta_b}(k, \eta) }
	\,,
\nonumber\\
\eal {4 \over 3} \left( { G \over \dot G} {k \over H_0} \right)^2 
{v_{\rm rms}^2 \over \Delta^2_{\delta_b}}
\label{eqn:Iap}
\end{eqnarray}

In Fig.~\ref{fig:I}, we show that for the $\Lambda$CDM model,  the density
and velocity contributions decouple in this manner beyond $k \sim 0.2 h$Mpc$^{-1}$.
The approximation (\ref{eqn:Iap}) represents an efficient way of 
evaluating the otherwise computationally expensive 
the mode coupling integral in this regime.

\subsection{Nonlinear Perturbations: Kinetic SZ Effect}
\label{sec:nonlinearV}

Nonlinear density fluctuations caught up in a bulk flow from larger
scales gives rise to the 
kinetic Sunyaev-Zel'dovich effect from large-scale structure.
It can alternately be thought of as the nonlinear extension of
the Vishniac effect.  
The reason is that in adiabatic CDM 
cosmologies nonlinearities only affect the density field 
below the coherence scale of the bulk velocity.
Recall that in this limit the Vishniac effect arises from
density perturbations caught in a large-scale flow to which
it is uncorrelated.

Figure \ref{fig:I} illustrates this fact for the fiducial
$\Lambda$CDM. 
Notice that even at
$z=0$, the dark matter density field only deviates substantially
from the linear 
approximation after equation~(\ref{eqn:Iap}) becomes a good
description of the Vishniac effect.   If we then
replace the linear {\it density} power spectrum with
its nonlinear analogue  but leave the contribution from the velocity
power spectrum the same, we obtain
\begin{eqnarray}
C_\ell^{\Theta\Theta} \eal {\pi^2 \over 2\ell^5 } \int
{d D}\, D_A^3 
\left(g {\dot G \over G} \right)^2 
%\nonumber\\ 
%\al\quad \times
\Delta^{2\, ({\rm lin})}_{\delta_b} \Delta^2_{\delta_b} I_V \,,
\label{eqn:vishnl}
\end{eqnarray}
with the mode coupling integral $I_V$ using 
equation~(\ref{eqn:modecoupling})
evaluated under linear theory.  
This expression has the nice feature that it expresses the total
effect: it includes both the
Vishniac effect and the kinetic SZ effect from nonlinear structures.

The underlying assumption is that the density fluctuations 
in the nonlinear regime are uncorrelated with the bulk velocity field.
Nonlinear evolution in 
the density field will correlate velocity and density modes if they
are both in the nonlinear regime.  
However, the bulk flow in adiabatic CDM models
arises mainly from the linear regime.  In the $\Lambda$CDM model,
half the contributions to $v_{\rm rms}^2$ comes from scales 
$k<0.07h $Mpc$^{-1}$ and the fluctuations go nonlinear at
$k \sim 0.2 h$Mpc$^{-1}$. 
\cite{ScoZalHui99} (1999) 
found that for the 4-point statistics, nonlinear modes are highly
correlated but that the correlation drops rapidly as 
one pair of the modes enters the linear regime.  
Decorrelation is even more effective here since
the relevant velocity and density modes are oriented perpendicular to 
each other.

Once the nonlinear power spectrum of the {\it baryonic gas} is known,
the resultant anisotropies can be calculated with equation~(\ref{eqn:vishnl}).
This unfortunately requires hydrodynamic simulations in general and {\it ab initio}
attempts in the literature (e.g. \cite{Peretal95}\ 1995) to simulate
the scattering effects have not had the dynamic range to treat the 
full problem.  Let us instead break the
problem into two: calculate the effect under the assumption that
the gas tracks the dark matter and then estimate where this 
approximation may break down.

\begin{figure}[t]
\centerline{
\epsfxsize=3.5truein\epsffile{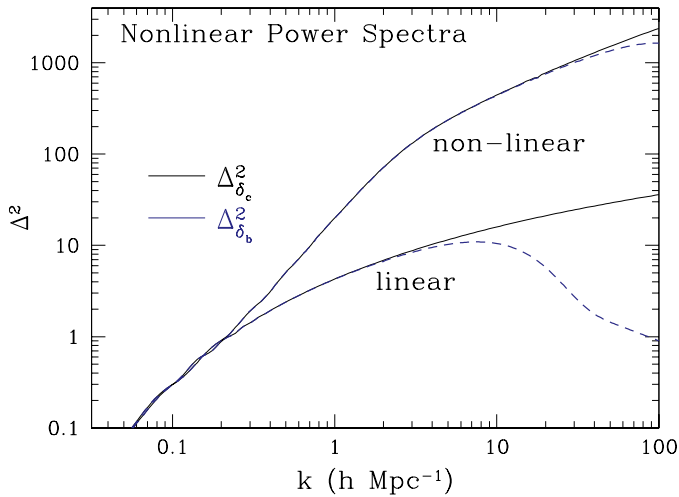}}
\caption{Nonlinear vs. linear power spectra for the CDM under the \protect\cite{PeaDod96} (1996)
scaling approximation and for the baryons under the additional 
prefiltering ansatz
of \protect\cite{GneHui98} (1998).
Note that the dilation of scales brings the filtering scale deep into the nonlinear regime.}
\label{fig:density}
\end{figure}

\subsubsection{Maximal Effect}

Let us first consider the effect under the simple assumption that
the gas traces the dark matter to place an upper limit on the magnitude of the
effect.   
The nonlinear dark matter power spectrum has been well-studied with
N-body simulations for the full range of adiabatic CDM cosmologies.
\cite{Hametal91} (1991) suggested a useful scaling relation for the
correlation function in the nonlinear regime that was generalized
to the power spectrum by \cite{PeaDod94} (1994).  The underlying
idea is that nonlinear fluctuations on a scale $k$ arise from linear
fluctuations on a larger scale
\begin{equation}
k_{\rm lin} = [1+\Delta_{\delta_c}^2(k)]^{-1/3} k \,,
\end{equation}
so that there is a functional relation between the nonlinear and linear power
spectra at these two scales
\begin{equation}
\Delta_{\delta_c}^{2}(k) = f_{\rm NL}[\Delta_{\delta_c}^{2\, {\rm (lin)}}
	(k_{\rm lin})]\,,
\label{eqn:PDscaling}
\end{equation}
which may be fit to simulations.  We take the \cite{PeaDod96} (1996;
eqn. [21]-[27]) form for $f_{\rm NL}$.  The resulting power spectrum for the
$\Lambda$CDM model are shown in Fig.~\ref{fig:density}.

\begin{figure}[t]
\centerline{\epsfxsize=3.5truein\epsffile{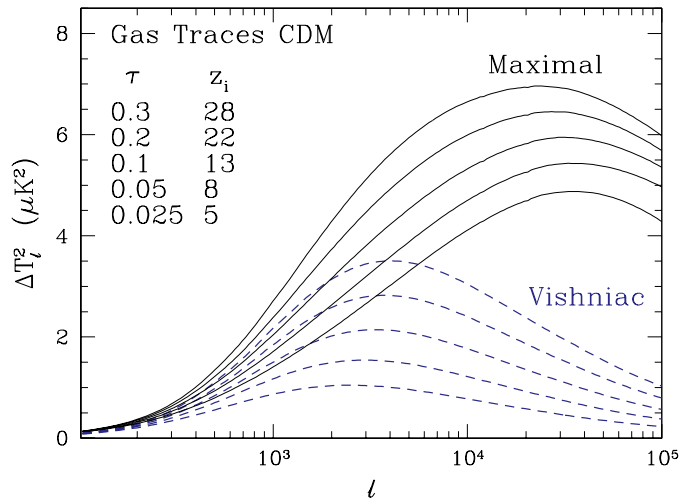}}
\caption{Maximal nonlinear enhancement of the Vishniac effect.  Under the assumption that the gas density traces the
dark matter density into the deeply nonlinear regime the Vishniac effect is significantly enhanced by
nonlinearities at $\ell \ga 1000$ especially in the late reionization scenarios.}
\label{fig:nonlin}
\end{figure}

Under this assumption, the kinetic SZ effect provides a
significant enhancement of the Vishniac effect.  As shown in Fig.~\ref{fig:nonlin},
the relative contribution
is the largest for late reionization scenarios (low $z_i$) because anisotropies
then arise from structure that is well-developed.    
However one must keep in mind that this is essentially an upper limit to the contributions since
the gas pressure will smooth out the gas density below the Jeans scale.

\begin{figure}[t]
\centerline{ \epsfxsize=3.5truein\epsffile{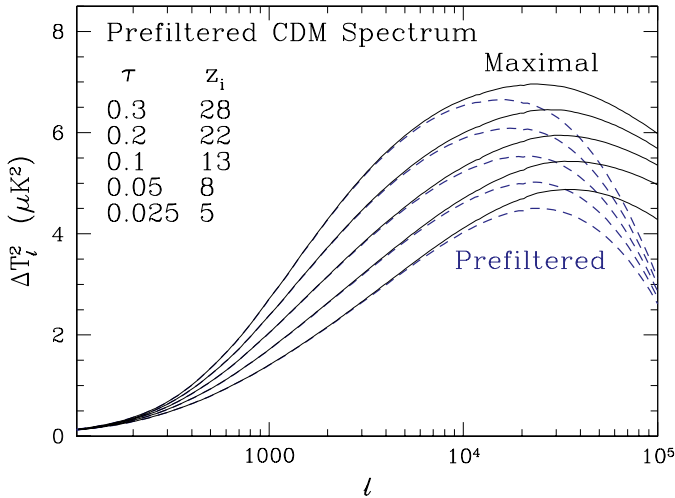}}
\caption{Prefiltering the dark matter spectrum.  Prefiltering the power spectrum to account for
gas pressure as in Fig.~\protect\ref{fig:density} cuts off the anisotropies at $\ell \sim 10^4$ for
the $\Lambda$CDM model.}
\label{fig:filt}
\end{figure}

\subsubsection{Pressure Cutoff}

\cite{GneHui98} (1998) examined simple schemes to approximate the
effect of gas pressure.
One such scheme that has fractional errors on the 10\% level for
overdensities $\la 10$ is to filter the density perturbations in
the linear regime as $\Delta_{\delta_b}^2 = f_b^2(k/k_{\rm F})
\Delta_{\delta_c}^2$ and treat the system as collisionless baryonic
particles.  Their best fit is obtained with the filter
\begin{equation}
f_b = {1 \over 2} [ e^{-x^2} + {1 \over (1+4 x^2)^{1/4}}]
\end{equation}
and \cite{Gne98} (1998) suggests $k_{\rm F} = 34 \Omega_m^{1/2} h$Mpc$^{-1}$
as a reasonable choice for the thermal history dependent filtering scale.

The results of applying the nonlinear scaling relation of 
equation~(\ref{eqn:PDscaling}) to the filtered linear power spectrum are shown in Fig.~\ref{fig:density}.
Notice that the dilation in wavenumber takes the filtering scale deep
into the nonlinear regime.  The calculations should not be trusted beyond
this scale since the nonlinear scaling breaks down for spectra with
a sharp cut off; indeed we have taken the local slope of the unfiltered
spectrum rather than the filtered spectrum when evaluating the \cite{PeaDod96}
(1996) formulae.  Still the result displayed in Fig.~\ref{fig:filt} 
imply that the gas-traces-CDM assumption is reasonable in the
arcminute regime $\ell \la 10^4$.  

\begin{figure}[t]
\centerline{\epsfxsize=3.5truein\epsffile{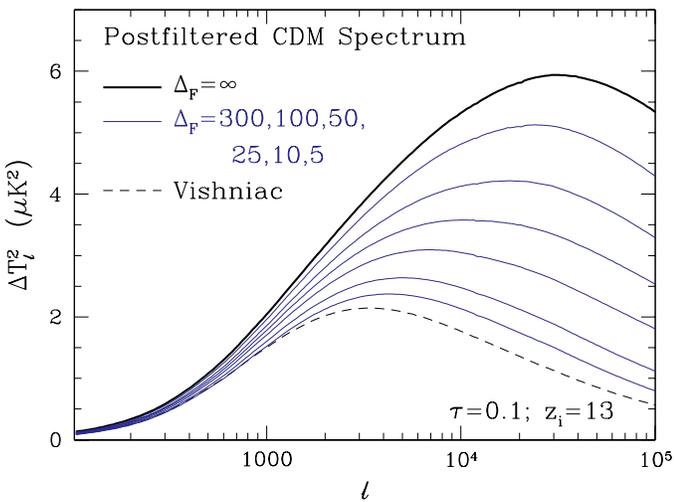}}
\caption{Postfiltering the dark matter spectrum.  
Filtering the CDM power spectrum 
with equation~(\protect\ref{eqn:smooth}) 
introduces a smooth
cut off at small angles 
and shows that nonlinearities are important as long as the gas traces the dark matter out to $\Delta_F \sim 10$.}
\label{fig:smooth}
\end{figure}

To give a more model-independent quantification of this effect, let us  also
consider a baryonic power spectrum given by
\begin{equation}
\Delta_{\delta_b}^2(k) = \Delta_{\delta_c}^2(k) 
	\exp(-\Delta_{\delta_c}^2/\Delta_{F}^2) \,.
\label{eqn:smooth}
\end{equation}
The results for several values of $\Delta_{F}$ are given in 
Fig.~\ref{fig:smooth} and show that only if the baryons fail to trace
the dark matter out to $\Delta_{F} \sim 10$ can nonlinear effects
be ignored. 

Our analysis shows that the kinetic SZ effect should be 
an important contributor to small-angle anisotropies.  
On arcminute scales and above, the assumption that the gas traces
the dark matter is reasonable given our current understanding
of the thermal history of adiabatic CDM models.
The amplitude of the effect at the subarcminute scales will depend 
on the details of thermal history throughits
effect on the clustering of the gas.  
Observations in this regime may open a new window
on the physics of this regime. Isolation of this effect from foregrounds (e.g. radio and infrared point sources)
and other secondary anisotropies (e.g. the inhomogeneous reionization signal considered in \S \ref{sec:patchy}) 
will be the main obstacle to overcome. 

\begin{figure}[t]
\centerline{\epsfxsize=3.5truein\epsffile{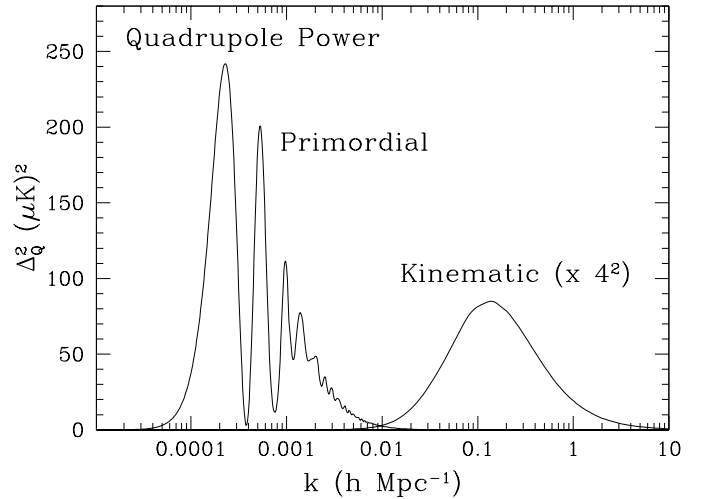}}
\caption{The power spectrum of the present-day quadrupole from the projection of primordial temperature inhomogeneities
and the kinematic Doppler effect in the $\Lambda$CDM model.  
The primordial quadrupole has been calculated numerically with a Boltzmann code.}
\label{fig:pquad}
\end{figure}

\section{Polarization from Homogeneous Scattering}
\label{sec:homopol}

In this section we consider the origin of quadrupole anisotropies and the 
linear polarization
they generate through Thomson scattering in a homogeneous medium.  The quadrupole anisotropies fall into
three broad classes: the primordial quadrupole from the projection of Sachs-Wolfe temperature anisotropies,
the intrinsic quadrupole from scattering (\S \ref{sec:primint}), and the kinematic quadrupole from
the second order Doppler effect (\S \ref{sec:kinematicquad}).

\subsection{Primordial and Intrinsic Quadrupoles}
\label{sec:primint}

A large-scale temperature inhomogeneity at the recombination epoch ($z \sim 10^3$) from
the Sachs-Wolfe gravitational redshift effect $\Theta=-\Phi/3$ becomes
a quadrupole anisotropy at low redshifts by simple projection.  In other words, the spherical
harmonic decomposition of the plane wave fluctuation,
\begin{equation}
-{\Phi \over 3}\exp (i {\bf k} \cdot {\bf x}) \rightarrow -{\Phi \over 3}j_\ell({k\over H_0} D) Y_\ell^0 \,,
\label{eqn:planedecomp}
\end{equation}
yields a spectrum of anisotropies and in particular a quadrupole
anisotropy given by
\begin{equation}
Q^{(0)} = -\sqrt{5} {\Phi \over 3} j_2({k\over H_0} D_*) \,.
\end{equation}
Here $D_* = D_{\rm rec}-D$ and we have assumed that spatial curvature can be neglected
at reionization.  Curvature corrections may be included by replacing the spherical bessel
function $j_2$ with the hyperspherical bessel function $\Phi_2^\nu$ (see Appendix).  

The logarithmic power spectrum of the quadrupole can 
therefore be expressed in terms of that
of the potential as
\begin{eqnarray}
\Delta_Q^{2\,(0)} \eal {5 \over 9} 
		j_2^2({k \over H_0} D_*)
		\Delta_\Phi^2 
\nonumber\\
	       \eal {5 \over 4} \delta_H^2 \left({ k \over H_0 }\right)^{n-1} (1+z)^2 G^2\Omega_m^2 j_2^2({k \over H_0} D_*)\,,
\end{eqnarray}
where the second line employs equation~(\ref{eqn:normalization})
and again assumes matter domination at $D$.
We show the results of numerically calculating the quadrupole power spectrum from a  
cosmological Boltzmann code
(\cite{WhiSco96} 1996) in Fig.~\ref{fig:pquad}.  Note that the spectrum peaks at large scales, corresponding to 
the horizon at the reionization epoch $k \sim H_0/D_*$.  On smaller scales, the projection 
(\ref{eqn:planedecomp}) carries the power to higher angular moments.

The rms is the integral over the power spectrum
\begin{eqnarray}
Q_{\rm rms}^2 \eal \int {d k \over k} \Delta_Q^{2\,(0)}  \,,
\nonumber\\
	      \eal {5 \over 4} \delta_H^2 (1+z)^2 G^2 \Omega_m^2 (D_* H_0)^{1-n} \int dx x^{n} j_2^2(x) \,,
\nonumber\\
	      \eal {5 \over 48} \delta_H^2 (1+z)^2  G^2 \Omega_m^2 (D_* H_0)^{1-n} \Gamma_{\rm SW}(n)\,,
\label{eqn:primordialQ}
\end{eqnarray}
where
\begin{equation}	
\Gamma_{\rm SW}(n)= 3\sqrt{\pi}
		{\Gamma[(3-n)/2] \over \Gamma[(4-n)/2]}  	
		{\Gamma[(3+n)/2] \over \Gamma[(9-n)/2]} \,. 	
\end{equation}
For reference, note that
$Q_{\rm rms}^2=5C_2^{\Theta\Theta}/4\pi$.
For our fiducial $\Lambda$CDM model $\delta_H=4.2 \times 10^{-5}$,  $(1+z) G = 1.24$, and 
$\Gamma_{\rm SW}(1) = 1$, we obtain $Q_{\rm rms} T_{\rm CMB}= 16 \mu$K.   

There are
two effects that modify this result slightly.  The first is that the universe may not
be completely matter-dominated at the time of scattering.  Additional anisotropies are 
then created as the gravitational potential decays.  
At high redshifts, the effect of
the radiation energy density pushes the quadrupole anisotropy up via the ``early'' integrated
Sachs-Wolfe effect (\cite{HuSug95} 1995); correspondingly at low redshifts the effect of
the cosmological constant or curvature again pushes the quadrupole up via the ``late''
integrated Sachs-Wolfe effect (\cite{KofSta85} 1985).  These effects are clearly visible in
Fig.~\ref{fig:tquad}, where we plot
the evolution of the rms quadrupole from numerical
calculations.  

The second effect is that scattering itself
can produce quadrupole anisotropies through projection of the Doppler effect and also destroy the primordial
quadrupole through isotropization.  
We call the net effect the contributions of the intrinsic quadrupole.
This effect is suppressed both by the assumed optically thin conditions and by the fact 
that the Doppler effect is cancelled except on scales near the horizon where the velocity itself
is small.   In Fig.~\ref{fig:tquad}, this effect is barely visible as a blip near the redshift of
reionization.

The hallmark of all these effects is that they create only $m=0$ type quadrupoles, because they come
from ($\ell=0$, $m=0$)
gravitational potential perturbations and 
($\ell=1$, $m=0$) potential flows. They therefore generate 
only $E$-parity polarization.  
The polarization from these sources are not slowly varying across a wavelength of the horizon-sized
perturbations and hence the approximation developed in the Appendix cannot be used to calculate
these effects.  
They are however automatically included in cosmological Boltzmann codes.
We show the resulting polarization in Fig.~\ref{fig:pol} for the $\Lambda$CDM
model as calculated with CMBFast.  As is well known, the secondary polarization from these effects are
confined to the lowest $\ell$ since the quadrupole sources contribute mainly on horizon scales at
last scattering.  

\subsection{Kinematic Quadrupole}
\label{sec:kinematicquad}

As pointed out by \cite{SunZel80} (1980), in the rest frame of the
scatterers, an isotropic CMB gains a quadrupole anisotropy 
from the quadratic Doppler effect
\begin{equation}
\Theta  = {\sqrt{1-v_b^2} \over 1 - \hat{\bf n} \cdot {\bf v}_b} - 1
	\approx \hat{\bf n} \cdot {\bf v}_b 
	+ (\hat{\bf n} \cdot {\bf v}_b)^2 - {1 \over 2} v_b^2 \,,
\end{equation}
that induces a polarization in the CMB by Thomson scattering.  
In real space, the quadrupole moment is simply 
\begin{equation}
Q^{(0)}({\bf x}) = {2 \over 3\sqrt{5}} v_b^2({\bf x})
\label{eqn:Qreal}
\end{equation}
in the basis aligned with the velocity $\hat{\bf e}_3 \parallel {\bf v}_b
({\bf x})$;
in Fourier space with a basis $\hat{\bf e}_3 \parallel \hat{\bf k}$,
the expression is more involved
\begin{eqnarray}
Q^{(m)} \eal - \int d\Omega\, {Y_2^{m *}(\hat{\bf n})
\over \sqrt{4\pi}}
	    \int {d^3 k_1 \over (2\pi)^3} 
	\hat{\bf n} \cdot {\bf v}_b({\bf k}_1)\,
	\hat{\bf n} \cdot {\bf v}_b({\bf k}_2)
\nonumber\\
	\eal -{2\sqrt{4\pi} \over 15} 	
	    \int {d^3 k_1 \over (2\pi)^3} 
       	    v_b(k_1) v_b(k_2){ k_1 \over  k_2}
\nonumber\\
	\al \quad \times 
	    \left[ Y_2^{m*}(\hat{\bf k}_1) 
          - \sqrt{5 \over 3 + |m|} {k \over k_1} Y_1^{m*}(\hat{\bf k}_1) \right]\,,
\end{eqnarray}
where $Y_1^{2}=Y_1^{-2}=0$.
The logarithmic power spectrum becomes
\begin{equation}
\Delta_Q^{2\, (m)} = 
		{4 \over 45}
		\Delta_{v_b}^{2}(k) 
		\Delta_{v_b}^{2}(k)
		I_Q^{(m)} \,,
\label{eqn:kinematicquadm}
\end{equation}
where
the mode coupling integrals are given by
\begin{eqnarray}
I_Q^{(m)} \eal 	
		\int_0^\infty dy_1 
		\int_{-1}^1 d\mu 
		{a^{(m)} \over y_1  y_2^3} 
%\nonumber\\
%\al \times 	
		{\Delta_{v_b}^{2}(ky_1)  \over
		\Delta_{v_b}^{2}(k) }
		{\Delta_{v_b}^{2}(ky_2)
		\over 
		\Delta_{v_b}^{2}(k)} \,.
\end{eqnarray}
Here the angular arguments are 
\begin{eqnarray}
a^{(\pm 2)} \eal {3 \over 8}y_1^2 (1-\mu^2)^2 \,,
\nonumber\\
a^{(\pm 1)} \eal {3 \over 8}(1-2\mu y_1)^2 (1-\mu^2) \,,
\nonumber\\
a^{(0)} \eal {1 \over 4} [2\mu - y_1(3\mu^2-1)]^2\,,
\end{eqnarray}
and it is useful to define the auxilary quantity
\begin{equation}
a^{(f)} = {1 \over 4} (1-\mu^2)\,.
\end{equation}

\begin{figure}[t]
\centerline{\epsfxsize=3.5truein\epsffile{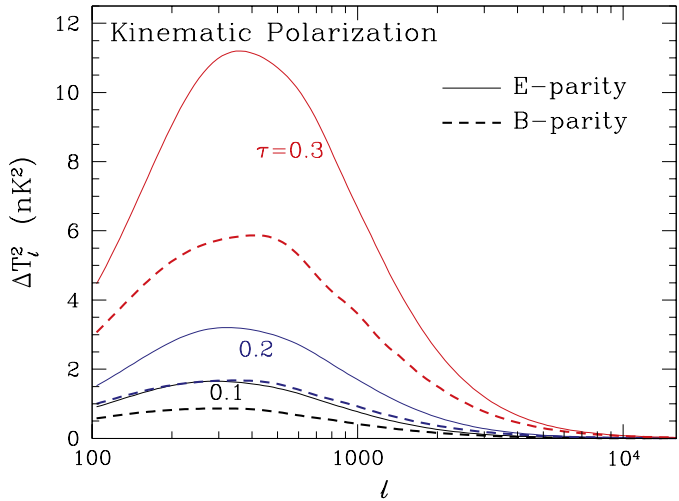}}
\caption{Power spectra of $E$- and $B$-parity polarization from the kinematic quadrupole in the $\Lambda$CDM model.}
\label{fig:kinpol}
\end{figure}

The polarization power spectrum then follows by
inserting equation~(\ref{eqn:kinematicquadm}) in
equation~(\ref{eqn:eebb}).  The result for the $\Lambda$CDM
cosmology is shown in Fig.~\ref{fig:kinpol}

The rms quadrupole follows simply from the relation
\begin{equation}
\sum_{m=-2}^2  a^{(m)}= y_2^2 - a^{(f)}
\end{equation}
with a change of variables to ${\bf k}_2$ for
the integration of the first term
\begin{eqnarray}
Q_{\rm rms}^2 \eal \int {d k \over k }\sum_{m=-2}^{2} \Delta_Q^{2\,(m)} 
= 
{8 \over 45}(1-f_{\rm kin}) v_{\rm rms}^4
\label{eqn:Qrmskinetic} 
\\
f_{\rm kin} \eal  
	{1 \over 2} \int {dk \over k}
		{\Delta_{v_b}^{2} \Delta_{v_b}^{2} \over v_{\rm rms}^4}
		I_Q^{(f)} \,.
\nonumber
\end{eqnarray}
For the $\Lambda$CDM model $f_{\rm kin} = 0.16$.
This expression is smaller than the naive expectation implied by 
real space relation (\ref{eqn:Qreal}),  
$4 \left<v_b^4\right>/45
= 12 v_{\rm rms}^4/45$, 
because the quadrupole is oriented 
and not simply a scalar gaussian random field.  
The main effect can
be understood as the contribution from the variance 
of $v_b^2$, $\left< v_b^4 \right>
- \left< v_b^2 \right>^2 = 2 v_{\rm rms}^4$, since its mean value 
comes from contributions that cancel out in orientation for the
quadrupole. 

\begin{figure}[t]
\centerline{\epsfxsize=3.5truein\epsffile{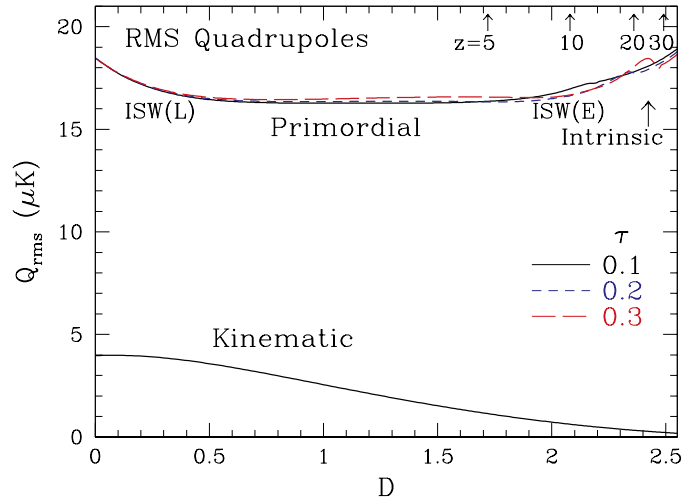}}
\caption{The time evolution of the rms quadrupole in the $\Lambda$CDM model.  Upper curves: the various linear theory effects
(see text) for three different ionization histories
from numerical calculations.  Lower curves: the kinematic effect.}
\label{fig:tquad}
\end{figure}

The kinematic quadrupole is not negligible in comparison with the primordial quadrupole for $\Lambda$CDM or
indeed any model that fits the observations. Large angle anisotropies are observed today at $10^{-5}$ 
while the velocity field reaches $10^{-3}$.  
As shown in Fig.~\ref{fig:tquad}, for $\Lambda$CDM it is approximately $1/4$ the primordial
one in amplitude today.  Furthermore, equation (\ref{eqn:kinematicquadm}) implies that it generates
comparable power in all 5 $m$-modes of the quadrupole and hence comparable $E$ and $B$ parity
polarization.   One might naively assume that the kinematic effect generates as much $B$-parity
polarization as the primordial quadrupole generates $E$.  This would then be an obstacle for detecting
gravity waves through the $B$-parity polarization.

There are however two reasons why the polarization is much smaller than implied by these arguments.
The first is that the $v_{\rm rms}^2$ declines as $(1+z)^{-1}$ in the matter-dominated epoch whereas the
primordial quadrupole remains constant.  The second is that 
in the $\Lambda$CDM model the coherence of the velocity field is over 2 orders of magnitude smaller
than that of the primordial quadrupole and hence generates polarization at $\ell \sim 400$.  Recall that 
the rule of thumb for projected sources is that the angular power spectrum is down by one factor
of $\ell$ from the spatial power spectrum.  This produces another order of magnitude suppression of
the polarization (see Fig.~\ref{fig:kinpol}).  The $B$-parity
polarization generated here
is significantly smaller than that generated by gravitational lensing of the primary polarization
(\cite{ZalSel98} 1998).  This effect is therefore unlikely to be detectable in $\Lambda$CDM models.

It is worth emphasizing that the amplitude of this effect is highly model dependent and scales in
power as $v_{\rm rms}^4 \lambda_v$ where $\lambda_v$ is the coherence scale of the velocity 
field.  Upper limits on the $B$-polarization can therefore constrain the amplitude and coherence of
the velocity field.  
For example consider a velocity field with an rms of $\sim 700 $km s$^{-1}$ on the a tophat scale of
150$h^{-1}$ Mpc is suggested by the data of \cite{LauPos94} (1994).  If we model that field
by the $\Lambda$CDM velocity power spectrum but increase the coherence length by a factor of
10 and the rms velocity by 1.55, we boost the polarization 
power by $1.55^4 \times 10 = 58$ and angular scale by 10 ($\ell \sim 40$).  The power would be
even larger in an open universe where the velocity amplitude actually grows substantially
with redshift before declining.
It is also worth bearing in mind that effects such as these 
warn against blindly taking even a large-angle detection of $B$-polarization 
as a {\it model-independent} detection of gravity waves or vorticity.

Finally, note that the correlation between primary and secondary temperature anisotropies
and the  kinematic polarization is suppressed since it involves the three-point function of the density field 
and hence vanishes in the linear regime.  

\section{Density-Modulated Polarization}
\label{sec:vishniacpol}

Here we consider the polarization generated from the 
density-modulation of the quadrupole anisotropies considered in the
last section.  Again these are separated by the nature of the quadrupole
source: the primordial quadrupole \S \ref{sec:modprim}, the kinematic
quadrupole \S \ref{sec:modkin}, and the intrinsic quadrupole \S \ref{sec:modint}.

\subsection{Primordial Quadrupole}
\label{sec:modprim}

\begin{figure}[t]
\centerline{
\epsfxsize=3.5truein\epsffile{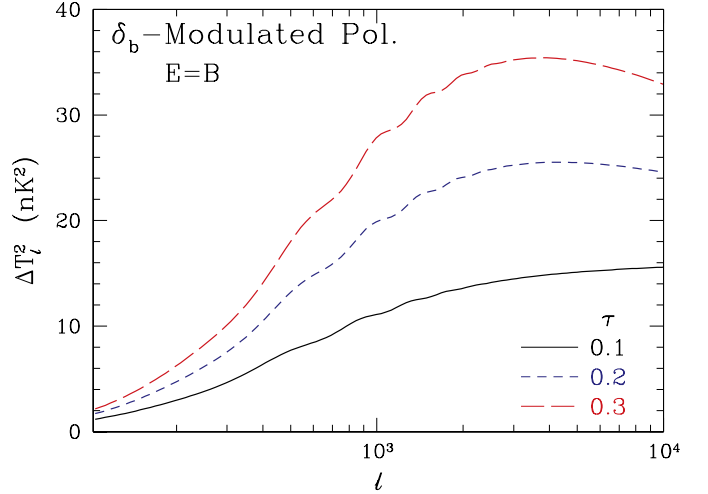} }
\caption{Density-modulated polarization in each parity type $E=B$ for the $\Lambda$CDM model. Here we assume that the gas traces the dark matter.}
\label{fig:poldensity}
\end{figure}

For the modulation of the primordial quadrupole by density 
fluctuations, we have
\begin{equation}  
Q^{(m)}_g (k) = \int {d^3 k_1 \over (2\pi)^3} Q^{(m)}({\bf k}_1)
	\delta_b(k_2)\,,
\end{equation}
Scalar perturbations in linear theory only generate $m=0$ quadrupoles in
the ${\bf k}_1$-basis, i.e. $Q^{(0)}(k_1) Y_2^0(\beta,\alpha)$ 
where $\beta$ and $\alpha$ are the polar and azimuthal angles 
defining $\hat n$ in this basis.  
One can project this onto the ${\bf e}_3 \parallel
{\bf k}$ basis with the help of the angular 
addition relation (see \cite{HuWhi97} 1997 eq. [7]),
\begin{equation} 
Y_2^0(\beta,\alpha) = \sqrt{4\pi \over 5} \sum_m Y_2^{m*}({\bf k}_1)
					    Y_2^m(\hat{\bf n})
\end{equation}
such that
\begin{equation}  
Q^{(m)}_g (k) = \sqrt{4 \pi \over 5} 
	\int {d^3 k_1 \over (2\pi)^3} 
	Q^{(0)}(k_1) Y_2^{m*}(\hat{\bf k}_1)
	\delta_b(k_2)\,,
\label{eqn:quadsource}
\end{equation}
Since the quadrupole source $Q^{(0)}(k_1)$ peaks on the scale of the
horizon whereas the density perturbations rise to small scales in
adiabatic CDM models, the correlation decouples
\begin{equation}  
\Delta_{Q_g}^{2\, (m)} = {1 \over 5}
	\Delta_{\delta_b}^2 Q^{2}_{\rm rms} \,,
\label{eqn:qcorr}
\end{equation}
where we have again used the orthogonality of spherical harmonics
and $Q^2_{\rm rms}$ is taken from numerical calculations as in 
Fig.~\ref{fig:tquad}.
Inserting equation~(\ref{eqn:qcorr}) into equation (\ref{eqn:eebb}),
we obtain
\begin{eqnarray}
C_\ell^{EE} = C_\ell^{BB} = { 3\pi^2 \over 50\ell^{3}} 
	\int dD\, D_A g^2 \Delta_{\delta_b}^2 
			Q_{\rm rms}^2 \,.
\label{eqn:clpolvish}
\end{eqnarray}
As shown in Fig.~\ref{fig:poldensity}, even for the gas-traces-CDM assumption 
the polarization generated by this effect in a $\Lambda$CDM model is small, much smaller 
than the polarization generated by lensing (cf.~Fig.~\ref{fig:pol}).

The cross correlation between multipole moments of the Vishniac effect and 
the polarization $C_\ell^{\Theta E}$ vanishes when averaged over the sky
due to the different angular dependence of the underlying dipole and
quadrupole sources.  Mathematically, this can be seen in the
sources (\ref{eqn:quadsource}) and (\ref{eqn:vishsource}); the orthogonality of the
spherical harmonics guarantees that these terms will integrate to zero
when summing over the directions of each $k$-mode.

Note however that there will be a strong correlation between the amplitude of the polarization
and the amplitude of the temperature fluctuations and may provide a means of pulling out the signal
in models where it is somewhat stronger. 

\subsection{Kinematic Quadrupole}
\label{sec:modkin}

The kinematic quadrupole is also modulated by the density field.
Its effect is identical to the primordial quadrupole
except 
equation (\ref{eqn:Qrmskinetic}) for the kinetic quadrupole 
is used 
in equation (\ref{eqn:clpolvish}),
\begin{equation}
C_\ell^{EE} = C_\ell^{BB} = { 4\pi^2 \over 375\ell^{3}} 
	\int dD\, g^2 D_A \Delta_{\delta_b}^2 
			(1-f_{\rm kin})v_{\rm rms}^4 \,,
\end{equation}
where recall that $f_{\rm kin}$ is a small correction due to the
directionality of the quadrupole moments.
The ratio integrands  between the modulated Doppler and cosmic polarizations is
\begin{equation}
{8 \over 45}(1-f_{\rm kin}) {v^{4}_{\rm rms} \over Q_{\rm rms}^2 } \,,
\end{equation}
Since observationally $v_{\rm rms} \sim(1+z)^{-1/2}10^{-3}$ and $Q_{\rm rms} \sim 10^{-5}$, the effect
from the primordial quadrupole is typically larger.   We show the results for the fiducial
$\Lambda$CDM model in Fig.~\ref{fig:pol}.

\subsection{Intrinsic Quadrupole}
\label{sec:modint}

The quadrupole anisotropy generated by the Vishniac effect during
reionization produces a linear polarization intrinsic to the Vishniac
effect.  This effect is doubly suppressed by cancellation: 
both the  source of the quadrupole and its polarization 
contributions
along the line of sight cancel when integrating along the line of
sight.

The quadrupole moment generated by the Vishniac effect is given by
equation (\ref{eqn:xmoments}) 
\begin{eqnarray}
Q^{(m)} \eal \sqrt{5} \int dD \, g(D) v_g^{(m)} \alpha_{01,2}^{(m)}
\nonumber\\
	\aal \sqrt{3\pi^2 \over 20}
		{H_0 \over k}
	X \tau_H (1+z)^2 v_g^{(m)}  \,,
\end{eqnarray}
and thus the power per logarithmic interval
\begin{eqnarray}
\Delta_Q^{2\,(m)} \eal {3 \pi^2 \over 20}\left( {H_0 \over k} 
\right)^2 X^2 \tau_H^2 (1+z)^4 
\Delta_{v_g}^{2\,(m)} \,. 
\end{eqnarray}
The polarization power spectrum becomes
\begin{eqnarray}
C_\ell^{BB}\eal {9 \pi^4 \over 800 \ell^7} \int dD\, D_A^5
	X^2 \tau_H^2 (1+z)^4 \left( g {\dot G \over G} \right)^2 
\nonumber\\
	\al \quad\times \Delta_{\delta_b}^{2\,{\rm (lin)}} 
	\Delta_{\delta_ b}^2 I_V\,,
\end{eqnarray} 
with $C_\ell^{EE}=0$ since only $m=\pm 1$ contributes to the
Vishniac effect.
The integrand in this equation is down by a factor of 
\begin{equation}
{9\pi^2 \over 400 \ell^2} D_A^2 X^2 \tau_H^2 (1+z)^4 
= {\cal O}\left[{\tau^2 \over \ell^2} (1+z)\right] \,,
\end{equation}
compared with the temperature anisotropies and is thus highly suppressed 
at high $\ell \sim 10^3$ from an amplitude
that is already reduced by the low optical depth $\tau \la 10^{-1}$ 
in the reionized universe.
Finally, there is no cross correlation between the temperature and
intrinsic polarization  since the polarization is purely of $B$-parity.

\begin{figure}[t]
\centerline{\epsfxsize=3.5truein\epsffile{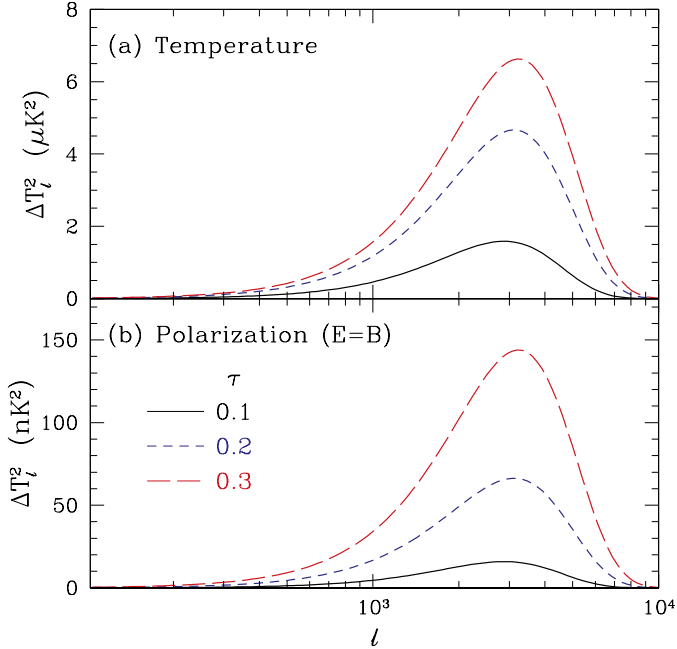}}
\caption{Ionization-modulation effect.  A patch size of $5$Mpc and duration of
patchiness of $\delta z_i /(1+z_i)=0.25$ in the $\Lambda$CDM model is
assumed.}
\label{fig:patchy}
\end{figure}

\section{Patchy Reionization}
\label{sec:patchy}

Inhomogeneities in the ionization fraction create a modulated Doppler
anisotropies and polarization of the CMB just as inhomogeneities in the density.
The techniques used in the calculation of density-modulated effects thus can be carried over
to here with little modification.  

For the modulated doppler effect, one takes ${\bf v}_g$ of 
equation~(\ref{eqn:vmodulated}) to be ${\bf v}_b \delta_{X}$
where $\delta_{X}$ is the fluctuation in the ionization fraction. 
Once the power spectrum of this quantity is known, the anisotropies
follow directly from equation~(\ref{eqn:dopplervorticity}).

Unfortunately, even the crude form of the power spectrum is not known
as it
requires not only an understanding of the first baryonic objects in the unvierse but also a full 
three dimensional radiative transfer in an inhomogeneous medium.  Thus even though we would expect
the ionizing sources to be associated with the peaks in the density field, it may be that the 
radiation escapes into and ionizes the low density medium where the recombination rates are the lowest 
(\cite{MirHaeRes99} 1999).
The techniques introduced here are therefore more useful for the inverse problem.  If such a signal
is detected in the CMB, we will want to use these relations to invert it to uncover the power spectrum
of the ionization fraction and hence learn about the manner in which the universe was ionized. 

Nevertheless it is useful to have a concrete model to illustrate these effects. 
\cite{GruHu98} (1998) introduced a toy model in which the universe is
reionized in uncorrelated spherical patches of a characteristic
size $R$ in a redshift interval
$\Delta z$ around $z_i$.   A distribution in $R$, like the one considered by \cite{Aghetal96} (1996), can of
course be constructed by superimposing simple sources.  
It is important to bear in mind that
 in a more realistic
scenario there are likely to be correlations between the ionized regions; 
\cite{KnoScoDod98} (1998) work out the consequences of ionized regions tracing the peaks in the density 
field.  
In the simple uncorrelated model, the correlation fuction of the ionization
fraction can be modeled 
as (\cite{GruHu98} 1998)
\begin{eqnarray}
\left< \delta X({\bf x}_1) \delta X({\bf x}_2) \right>\eal
	[ X - 
	  X^2]
%	\nonumber\\
%	\al \quad \times
	e^{-[({\bf x}_1 - {\bf x}_2)^2/2R^2]}\,,
\end{eqnarray}
where $X(D)$ now is the {\it mean} ionization. 
This form is chosen to have the right asymptotic behavior in time
and scaling with $R$; other forms can be chosen 
which essentially correspond to a redefinition of the patch size $R$
(e.g. the tophat spheres of \cite{KnoScoDod98} 1998)

With this ansatz, the logarithmic power spectrum becomes
\begin{equation}
{ X^2} \Delta_{\delta_{X}}^2 = {\sqrt{2 \over \pi}} (k R)^3 e^{-k^2 R^2/2}
	[ X -  X^2]\,.
\end{equation} 
The angular power spectrum that results is simply 
that of equation~(\ref{eqn:vishincoherent}) with the density power
spectrum replaced with the ionization power spectrum
\begin{eqnarray}
C_\ell^{\Theta\Theta} \eal {2\pi^2 \over 3\ell^3} \int
{dD}\, D_A g^2
\Delta_{\delta_{X}}^2 v_{\rm rms}^2 \,, 
\\
\eal {(2\pi)^{3/2} \over 3} \int dD\, D_A 
{g^2 \over X^2 }
\theta_0^3
e^{-\ell^2\theta_0^2/2 } [X-X^2] v_{\rm rms}^2 \,,
\nonumber
%\al\quad\times
%[ X -  X^2] 
\label{eqn:patch}
\end{eqnarray}
where $\theta_0 = H_0 R/D_A$.

Following \cite{GruHu98} (1998), we take a mean ionization fraction that
grows linearly from zero at $D_i$ up to unity by $D_i + \delta D_i$
\begin{equation}
X = {D - D_i \over \delta D_i} \,.
\end{equation}
Assuming reionization occurs promptly ($\delta D_i/D_i \ll 1$),
the other quantities may be taken out of the integral leaving
\begin{equation}
C_\ell^{\Theta\Theta} = {(2\pi)^{3/2} \over 18} 
\theta_0^3 {D_A \delta D_i}
{g^2 \over { X^2} }
e^{-{\ell^2\theta_0^2/ 2 } } 
v_{\rm rms}^2(D_i) \,,
\end{equation}

Likewise, the polarization follows from 
equation~(\ref{eqn:clpolvish})
\begin{eqnarray}
C_\ell^{EE} \eal C_\ell^{BB} = { 3\pi^2 \over 50\ell^{3}} 
	\int dD\, D_A \left({ g \over X} \right)^2 
		\Delta_{\delta_b}^2 Q_{\rm rms}^2 \,,
	\nonumber\\	
\eal { (2\pi)^{3/2} \over 200} 
	\theta_0^3 {D_A \delta D_i } {g^2 \over X^2}
	e^{-{\ell^2 \theta_0^2/ 2} } 
			Q_{\rm rms}^2(D_i) \,.
\end{eqnarray}
or
\begin{equation}
{C_\ell^{EE} \over C_\ell^{\Theta\Theta}} 
= {9 \over 100} {Q_{\rm rms}^2 \over v_{\rm rms}^2 }\Big|_{D_i} \,.
\end{equation}
with $Q_{\rm rms}$ being either the primordial quadrupole of
equation~(\ref{eqn:primordialQ}) or the kinematic quadrupole of equation~(\ref{eqn:Qrmskinetic}).
For adiabatic CDM models, the former typically dominates the latter even more so than for the density-modulated
effect.  The reason is that one no longer is weighted toward low redshifts and larger velocities by 
the growth of density perturbations. 
An example of the temperature and polarization signals for a specific choice of parameters
is given in Fig.~\ref{fig:patchy}.

\section{Discussion}

We have explored the role of mildly-nonlinear density fluctuations in
generating secondary CMB anisotropies and a host of contributions
to the secondary polarization.  In adiabatic CDM models, the Doppler effect
from scattering off nonlinear baryonic clumps in a large-scale bulk flow 
is a natural extension of the Vishniac effect.  In the small scale limit, the Vishniac effect 
simplifies to the corresponding effect for linear density perturbations.
For this class of models, the nonlinear contributions out to overdensities of
$\sim 10$ are comparable to or greater than the linear contributions, depending on the
redshift of reionization.  If the gas traces the dark matter to much higher overdensities
then upto an order of magnitude increase in the subarcminute anisotropy would result. 
Likewise small-scale patchiness in the ionization can greatly increase the anisotropy in this
regime.  Calculations of the clumpiness of the ionized gas are difficult and unreliable 
even with state-of-the-art numerical simulations.   An observational study of
subarcminute scale anisotropies therefore offers the opportunity to discover aspects of
reionization that are intractable to theoretical analysis today.

The secondary polarization from reionization is, as expected, extremely small in
the context of adiabatic CDM models for structure formation and unlikely to inhibit 
extraction of even subtle effects such as the $B$-parity polarization from gravitational
lensing or gravitational waves.  It is worthwhile to note that these mechanisms generically
predict comparable power in $E$ and $B$ parity polarization and may be 
important outside of
the adiabatic CDM model context.
For example, in a model with a coherent
$1500$km s$^{-1}$ velocity field on the $100 h^{-1}$Mpc scale the kinematically induced quadrupole
generates a $B$-parity polarization at the $0.1 \mu$K 
level for reasonable optical depths $\tau \sim 0.1-0.3$.  
Likewise, if the velocity field fell off less rapidly than $(1+z)^{-1/2}$ out to redshifts of order 10 or
more, then the effect might also be enhanced to this level.  In these models, the density and ionization
modulation of the scattering produce even larger signals in the arcminute regime that may be observable.
Thus even a null detection of $B$-parity polarization can be interesting.
The tools that we have introduced in the text and appendix are quite general and have applications beyond the 
adiabatic CDM model.

Even in the adiabatic CDM class of models, significant uncertainties remain expecially in the contribution from inhomogeneities in the ionization and
the small-scale behavior of the gas.  These questions will ultimately be answered by the observations themselves
as CMB experiments probe the arcminute regime with higher and higher sensitivity.

\appendix
\section{Generalizing the Limber Equation}

The \cite{Lim54} (1954) equation describes the two-point statistics
of a field which is the two-dimensional projection on the sky of
a three-dimensional source field whose statistical properties vary
slowly along the line of sight.   \cite{Kai92} (1992, 1998) expressed
the result as a relation between the two and three dimensional power
spectra of the fields.  \cite{HuWhi96} (1996) generalized these flat-sky
approximations in an all-sky approach for spatially flat cosmologies. 

Here we extend the techniques in three ways.  We generalize them to
open cosmologies where Fourier analysis on the three-dimensional field is invalid. 
We allow for
an arbitrary angular dependence in the source.  Finally,
we treat tensor fields on the sky in addition to the usual scalar fields. 

\subsection{Total Angular Momentum Method}

A general field $X$ that depends both on position ${\bf x}$ and direction 
${\bf n}$ at ${\bf x}$ 
can be expanded in a complete set of modes
denoted ${}_s G_j^m$ (see \cite{Huetal98} 1998), 
\begin{equation}
X({\bf x},{\bf n}) = \int {d^3 q \over (2\pi)^3} \sum_{\ell m} X_\ell^{(m)} {}_s G_\ell^m\,,
\end{equation}
where the 
spin index $s=0,\pm 2$ for scalar and tensor fields on the sky and
($\ell$,$m$) describe the multipole moments of the local angular dependence.
In flat space, these modes are simply the product of plane waves and
spin-spherical harmonics
(\cite{Goletal67} 1967)
\begin{eqnarray}
{}_s G_\ell^m &=& (-i)^\ell \sqrt{4\pi \over 2\ell+1} {}_s Y_\ell^m(\hat{\bf n})
	\exp(i {\bf q} \cdot {\bf x}) \,.
\end{eqnarray}
Note that ${}_0 Y_\ell^m = Y_\ell^m$, the ordinary spherical harmonics.
The angular power spectrum of the field is defined as
\begin{equation}
C_\ell^{XX} = 4\pi \int {d^3 q \over (2\pi)^3} \sum_m {X_\ell^{(m)*} X_\ell^{(m)} \over (2\ell+1)^2} \,.
\label{eqn:powerspectrumdef}
\end{equation} 

Let us suppose that the field on the sky $X$ is generated by the line-of-sight 
integral of another
positionally and directionally dependent field $S$ evaluated at a
${\bf x}=D\hat{\bf n}$
\begin{eqnarray}
X({\bf x},{\bf n}) \eal \int dD\, S({\bf x},{\bf n})  \,,
\\
 \eal \int dD\, \int{ d^3 q \over (2\pi)^3} \sum_{j m} S_j^{(m)} {}_s G_j^m \,.
\nonumber
\end{eqnarray}
The normal modes can be rewritten in spherical coordinates as
\begin{equation}
{}_s G_j^m =\sum_{\ell} (-i)^{\ell} \sqrt{4\pi(2\ell+1)}
			\alpha_{sj ,\ell}^{(m)}\,
			{}_s Y_{\ell}^m \,,
\label{eqn:normalmode}
\end{equation}
where we have used the spherical harmonic decomposition of a plane wave
to combine the local and plane-wave angular dependence terms into
the total angular dependence as seen by the observer at the origin.
Here the $\alpha_{s j ,\ell}^{(m)}$ are linear combinations of spherical
Bessel functions with weights given by Clebsch-Gordan coefficients as we shall see.
The moments of the field are then
\begin{eqnarray}
X_\ell^{(m)} = \int dD\, (2\ell+1) \sum_j S_j^{(m)} \alpha_{s j, \ell}^{(m)}\,.
\label{eqn:xmoments}
\end{eqnarray}

For open geometries, the expressions in spherical coordinates take the same form 
except the
radial functions 
$\alpha_{s j, \ell}^{(m)}$
are linear combinations 
of the hyperspherical Bessel
functions $\Phi_\ell^\nu$ (\cite{Huetal98} 1998).  
They depend separately on radial distance 
\begin{equation}
\chi = \Omega_K^{1/2} D \,,
\end{equation}
and wavenumber,
\begin{equation}
\nu = {q \over H_0 \Omega_K^{1/2}} \,,
\end{equation}
which itself differs from the usual eigenvalue $k$ of the Laplacian near 
the curvature scale
\begin{equation}
q = \sqrt{k^2 - (|m|+1)H_0^2\Omega_K} \,.
\end{equation}
Useful radial functions include 
\begin{eqnarray}
\alpha_{00,\ell}^{(0)}
        \eal \Phi_\ell^\nu \, , \nonumber\\
\alpha_{01,\ell}^{(1)}
\eal\sqrt{\ell(\ell+1)\over2(\nu^2+1)} {\rm csch}\chi\,\Phi_\ell^\nu\,,
        \nonumber\\
\alpha_{02,\ell}^{(2)}
\eal \sqrt{{3 \over 8}{(\ell+2)(\ell^2-1)\ell \over (\nu^2+4)(\nu^2+1)}}
        {\rm csch}^2\chi\, \Phi_\ell^\nu \,,
        \nonumber\\
\alpha_{01,\ell}^{(0)}
\eal \sqrt{1 \over \nu^2 + 1} \Phi_\ell^\nu{}' \, , \nonumber\\
\alpha_{02,\ell}^{(0)}
\eal {1 \over 2} \sqrt{1 \over (\nu^2+4)(\nu^2+1)}
\left[ 3 \Phi_\ell^\nu{}'' + (\nu^2+1)\Phi_\ell^\nu 
        \right] \, , \nonumber\\
\alpha_{02,\ell}^{(1)}
        \eal \sqrt{ {3 \over 2} {\ell(\ell+1) \over (\nu^2+4)(\nu^2+1) }}
           \left[ {\rm csch}\chi \Phi_\ell^\nu(\chi) \right]' \,.
\label{eqn:phiradial}
\end{eqnarray}
for $s=0$  and
\begin{eqnarray}
\alpha_{\pm 22,\ell}^{(0)}
\eal \sqrt{{3 \over 8}{(\ell+2) (\ell^2-1)\ell \over (\nu^2+4)(\nu^2+1)}}
        {\rm csch}^2\chi \Phi_\ell^\nu \,,
\nonumber
\\
\alpha_{\pm 22,\ell}^{(1)}
\eal {1 \over 2} \sqrt{(\ell-1)(\ell+2) \over (\nu^2 +4)(\nu^2+1)}
{\rm csch}\chi [({\rm coth}\chi \pm i\nu)\Phi_\ell^\nu
\nonumber\\
\al\quad+ \Phi_\ell^\nu{}']
\,,\nonumber\\
\alpha_{\pm 2 2,\ell}^{(2)}
        \eal {1 \over 4} \sqrt{1 \over (\nu^2+4)(\nu^2 + 1)}
[\Phi_\ell^\nu{}'' + 2( 2{\rm coth}\chi \pm i \nu )\Phi_\ell^\nu{}' 
\nonumber\\
\al\quad
+ ( 2 {\rm coth}^2 \chi+1 -\nu^2 \pm 4i \nu{\rm coth}\chi) \Phi_\ell^\nu]
 \,,
\label{eqn:epsilonradial}
\end{eqnarray}
for $s=2$.  Note that primes are derivatives with respect to $\chi$ and
$\alpha_{s j, \ell}^{(-m)} = \alpha_{s j,\ell}^{(m)*}$.

\subsection{Weak-Coupling Approximation}

The weak-coupling approximation was introduced in \cite{HuWhi96} (1996) as a means
of evaluating equation (\ref{eqn:xmoments}) in the limit that the source $S_j^{(m)}$
varies in distance $D$ only on a much larger scale than the wavelength $q^{-1}$.  
In this case, the source may be taken out of the integral and evaluated at
\begin{eqnarray}
D \eal \Omega_K^{-1/2} \sinh^{-1}(\Omega_K^{1/2} D_A) \,,
\nonumber\\
D_A \eal (\ell +{1 \over 2}) {H_0 \over q}\,.
\label{eqn:projectionrelation}
\end{eqnarray}
The remaining integral over $\alpha_{sj,\ell}^{(m)}$ becomes
\begin{eqnarray}
\int_0^\infty d D \, \alpha_{sj,\ell}^{(m)} \eal
{H_0 \over q} {\sqrt{\pi} \over 2} {\Gamma[(\ell+1)/2] \over 
		      \Gamma[(\ell+2)/2]} 
			W_{sj}^{(m)} 
			K_\nu 
\end{eqnarray} 
where the correction due to spatial curvature is given by
\begin{equation}
K_\nu = 
[1 + {(\ell+1/2)^2 \over \nu^2}]^{-1/4} \tanh^{1/2}({25 \over 16}\nu).
\end{equation}
and the weights $W_{sj}^{(m)}$ follow from the radial functions
(\ref{eqn:phiradial}) and (\ref{eqn:epsilonradial}):
\begin{eqnarray}
W_{00}^{(0)} \eal 1 \,, 
\nonumber\\
W_{01}^{(1)} \eal \sqrt{\ell(\ell+1) \over 2(\nu^2+1)}{\rm csch} \chi  \,,
\nonumber\\
W_{02}^{(2)} \eal \sqrt{{3 \over 8} {(\ell+2)(\ell^2 -1)\ell \over 
	(\nu^2 +4)(\nu^2+1)}} {\rm csch}^2\chi \,,
\nonumber\\
W_{01}^{(0)} \eal 0 \,,
\nonumber\\
W_{02}^{(0)} \eal {1 \over 2} \sqrt{(\nu^2+1) \over (\nu^2+4)} \,,
\nonumber\\
W_{02}^{(1)} \eal 0 \,,
\end{eqnarray}
for the spin zero fields and
\begin{eqnarray}
W_{\pm 22}^{(0)} \eal \sqrt{{3 \over 8} {(\ell+2)(\ell^2-1)\ell
	\over (\nu^2+4)(\nu^2+1)}} {\rm csch}^2 \chi    \,,
\nonumber\\
W_{\pm 22}^{(1)} \eal 
{1 \over 2} \sqrt{(\ell-1)(\ell+2)\over (\nu^2+4)(\nu^2+1)} {\rm csch}\chi
\nonumber\\
\al \quad\times ( \coth\chi \pm i\nu )\,, \\
W_{\pm 22}^{(2)} 
\eal {1 \over 4}\sqrt{1 \over (\nu^2+4)(\nu^2+1)}
\nonumber\\
\al \quad\times
\left[(1-\nu^2 + 2\coth^2\chi) \pm 4 i \nu \coth\chi\right] \,,
\nonumber
\end{eqnarray}
with
\begin{equation}
{\rm csch}\chi = \nu /(\ell+1/2)\,.
\end{equation}
Where there are zero entries in these equations, a
term of order $\dot S_j^{(m)} H_0/q$ exists and may be taken into account through
integration by parts.

Note that the real part is the contribution to the ``$E$''-parity 
of the tensor field and the imaginary part is the ``$B$''-parity contribution.
The flat limit is recovered as $\nu \rightarrow \infty$, $\chi \rightarrow 0$,
and $\nu\chi \rightarrow kD/H_0$.
We have numerically verified that the underlying expression
for $\int dD\, \Phi_\ell^\nu $ is good to 
$\sim 1\%$ for $\ell \ge 2$ using the code of \cite{Kos98} (1998).

\subsection{Limber Formulation}

The weak coupling approximation is 
a generalization of the familiar Limber equation
as noted by 
\cite{JafKam98} (1998).
In the weak coupling 
approximation, we are left with a power spectrum defined as the  integral over wavenumber 
$q$ in equation (\ref{eqn:powerspectrumdef}).  If we change variables to distance $D$
using the projection relation (\ref{eqn:projectionrelation})
we obtain
\begin{eqnarray}
C_\ell^{XX} 
\eal 2\pi^2 \int d D D_A F_\ell 
%\nonumber\\
%&&\quad
\sum_{m} |W_{sj}^{(m)}|^2 \Delta_S^{2\, (jm)}\,,
%\nonumber
\label{eqn:limberform}
\end{eqnarray}
for a source with a single type of angular dependence $S_j^{(m)}$ and
logarithmic power spectrum $\Delta_S^{2\, (jm)}$.
The weighting in $\ell$ is
\begin{eqnarray}
F_\ell \eal {1 \over 2(\ell+1/2)^{2}}
	\left[{\Gamma[(\ell+1)/2] \over \Gamma[(\ell+2)/2]}\right]^2
%\nonumber\\
%&&\times 
\tanh({25 \over 16}{\ell+1/2 \over \Omega_K^{1/2} D_A}) 
\nonumber\\
\aal {1 \over \ell^3} \,.
\end{eqnarray}
The approximation in the second line is for $\ell \gg 1$ where contributing
modes are well below the curvature scale for reasonable cosmologies
($\nu \gg 1$).
In this same limit, the weights also simplify,
\begin{equation}
\begin{array}{lll}
W_{00}^{(0)} = 1\,, \quad &
W_{01}^{(1)} = \sqrt{1 \over 2}\,, \quad &
W_{02}^{(2)} = \sqrt{3 \over 8}\,,
\\ 
W_{01}^{(0)} = 0\,, \quad &
W_{02}^{(0)} = {1 \over 2}\,, \quad &
W_{02}^{(1)} = 0\,, 
\\ 
W_{\pm 22}^{(0)} = \sqrt{3 \over 8}\,, \quad &
W_{\pm 22}^{(1)} = \pm {i \over 2}\,,  \quad &
W_{\pm 22}^{(2)} = -{1 \over 4}\,.
\end{array}
\end{equation}
Notice that for the spin-2 case, $m=0,\pm 2$ produces a 
pure ``$E$''-parity field and $m=\pm 1$ produces a pure ``$B$'' parity
field, in contrast to the tight-coupling (delta function source
approximation) where the power ratios between $B$ and $E$ are
$0,6,8/13$ for $m=0,\pm 1,\pm 2$ (\cite{HuWhi97} 1997).
For a simple source $S$ with no angular dependence ($j=0$, $m=0$) and
a scalar field ($s=0$) on the sky,
equation~(\ref{eqn:limberform}) reduces to
\begin{equation}
C_\ell^{XX} = 2{\pi^2 \over \ell^3} \int dD D_A {\Delta^2_S} \,,
\end{equation}
which 
is the Fourier Limber
equation as derived by \cite{Kai92} (1992; 1998).

{\it Acknowledgements:}  I would like to thank D.J. Eisenstein,
M. White, and M. Zaldarriaga for useful conversations.
W.H.\ is supported by the Keck Foundation, a Sloan Fellowship,
and NSF-9513835.

\end{document}